\documentclass[preprint,aps,pre,superscriptaddress]{revtex4-1}

\usepackage{latexsym,enumerate,amsmath,amssymb}
\usepackage{graphicx}
\usepackage{hyperref}
\usepackage{xcolor}
\usepackage{multirow}

\newcommand{\wt}[1]{\widetilde{#1}}

\def\bb{\mathbb}
\def\bm{\boldsymbol}

\newcommand{\comment}[1]{}

\begin{document}
\preprint{APS/123-QED}

\title{Parametric Reduced Models for the Nonlinear Schr\"odinger Equation}
\author{John Harlim}
\email{jharlim@psu.edu}
\affiliation{Department of Mathematics, the Pennsylvania State University, University Park, PA 16802-6400, USA.}%
\affiliation{Department of Meteorology, the Pennsylvania State University, University Park, PA 16802-5013, USA.}%
\author{ Xiantao Li}
\email{xli@math.psu.edu}
\affiliation{Department of Mathematics, the Pennsylvania State University, University Park, PA 16802-6400, USA.}%

\date{\today}

\begin{abstract}
Reduced models for the (defocusing) nonlinear Schr\"odinger equation are developed. In particular, 
we develop reduced models that only involve the low-frequency modes given noisy observations of these modes. The ansatz of the reduced parametric models
are obtained by employing a rational approximation and a colored noise approximation, respectively, on the memory terms and the random noise of a generalized Langevin equation that is derived from the standard Mori-Zwanzig formalism. 
The parameters in the resulting reduced models are inferred from noisy observations with a recently developed ensemble Kalman filter-based parameterization method. The forecasting skill across different temperature regimes are verified by comparing the moments up to order four, a two-time correlation function statistics, and marginal densities of the coarse-grained variables.
\end{abstract}

\pacs{05.10.-a,05.10.Gg,02.30.Zz}

\maketitle

\section{Introduction}
An important scientific problem in applied sciences is to forecast some quantity of interest of dynamical systems that exhibit multiscale behavior. Traditional approaches (see e.g., review paper \cite{GKS04}) often assume some knowledge about the underlying dynamics and proceed by deriving an effective equation for a set of preselected variables. Instead of  working with the trajectories associated with the full solutions, one is interested in a reduced model in which only the quantities  of interest are involved. These quantities of interest are generally referred to as the {\it coarse-grained} variables. In the case when the dynamics of the coarse-grained variables is of primary interest,  the effective model provides an efficient means to simulate directly the coarse-grained variables, without having to keep track of the remaining degrees of freedom.

An elegant framework for deriving the effective model is the Mori-Zwanzig projection \cite{Mori65,Zwanzig73,Zwanzig61}, 
which has recently become an extremely important tool to simplify complex dynamical systems \cite{ChKaKu98,espanol2004statistical,hijon2010mori,chorin2007problem,IzVo06,lange2006collective,oliva2000generalized,darve2009computing,venturi2014convolutionless,LiXian2014,stinis2012mori}. In particular, this derivation led to a set of generalized Langevin equation (GLEs), a typical result of the Mori-Zwanzig procedure. A notable feature of the GLE is a memory term which represents the history-dependence of the effective dynamics, along with a random noise term, which incorporates the influence of the remaining degrees of freedom. Unfortunately, solving the resulting GLE still remains as a challenge. For example, the memory function has been expressed as an infinite series \cite{chorin2000optimal}, and it may exhibit very slow decay. The implication is that a long history of the solution has to be kept in order to evaluate the integral in the GLE. The evaluation of the integral has to be done at every time step, which adds great complexity to the entire computation.  Furthermore, incorporating the random noise term is not straightforward.

A simple approach proposed by \cite{hijon2010mori,hijon2006markovian,kauzlaric2011bottom} is to approximate the memory kernel in the GLE model with a delta function (but with a carefully chosen damping parameter). This certainly introduces additional modeling error that is difficult to quantify.  In problems that arise from biological systems, the memory function in the GLE can be computed by matching the auto-correlation function of the coarse-grained variables. For instance, for the GLEs derived from Newton's equations of motion in classical mechanics, one can derive an integral equation for the memory function \cite{berkowitz1983generalized,berkowitz1981memory,lange2006collective}. But this approach requires the computation of the velocity correlation function for the {\it full model}, which clearly is a challenge.  Furthermore, the solution procedure for the integral equation is often not reliable.  Another approach to approximate the GLE is by using an extended Markovian system, which can be done using a projection to the Krylov subspace approximation \cite{LiXian2014,Li14}.  This approach, {however,} requires the knowledge of the full model, especially the interaction among all the degrees of freedom. But this approach suggests that the full GLE models with strong memory effects can be approximated by an extended system with a few auxiliary variables and this key result motivates the present work.

The main idea of the present work is to apply a rational approximation to the kernel function in the GLE and a colored noise approximation to the orthogonal dynamics in the GLE such that the resulting parametric model is Markovian. We subsequently use the stability conditions established in \cite{mh:13} as guidelines to ensure non blow-up solutions in the resulting models. Rather than deriving the explicit dependence of the parameters in the resulting Markovian models in terms of the true solutions (and/or the parameters in the original dynamics), we estimate these parameters by solving an inverse problem, filtering partially observed noisy measurement of the dynamics. This approach is often useful when (a) there is a large amount of training data, e.g., from experimental observation of part of the system; (b) the explicit form of the GLEs is difficult to obtain; (c) we don't have access to the exact solutions of the full dynamical systems. Computationally, since the resulting model is Markovian, we don't need to explicitly compute the memory terms and we don't need to store the solution history. More importantly, compared to direct numerical approximation of the integro-differential equations associated with the GLE model, solving the reduced parametric system requires much less computational cost.

We will demonstrate our modeling approach on the nonlinear Schr\"odinger equation (NLS), which finds many applications in various areas of applied physics. Of our particular interest is the statistical-mechanics aspects, which has been well studied theoretically \cite{bourgain1996invariant,lebowitz1988statistical,tzvetkov2008invariant}. Our goal is to predict the equilibrium statistical behavior of the low-order wave numbers. We should stress out that developing reduced models for the NLS equations is highly nontrivial in the following sense. Since the solutions of NLS equation exhibit strong correlation time with nontrivial autocorrelation function, the memory feedback from the unresolved scales is non-negligible and need to be appropriately accounted. Moreover, the equilibrium statistics of the solutions are highly non-Gaussian with bimodal distribution.
We will proceed by applying a rational approximation and a colored noise approximation, subsequently, to the kernel functions and orthogonal dynamics of a GLE, derived by Chorin and coworkers \cite{chorin2000optimal}. Subsequently, we estimate the parameters of the resulting model with an adaptive parameter estimation scheme that is recently developed in \cite{hmm:14}. We will then validate the forecasting skill by comparing moments up to order four, a two-time correlation function statistics, and marginal densities of the coarse-grained variables.  

The remaining part of the paper is organized as follows. In Section~\ref{sec2}, we state the problem and provide a short review of the GLE deduced by Chorin and coworkers~\cite{chorin2000optimal}. In Section~\ref{sec3}, we construct the parametric models. The procedure for the parametric estimation method in \cite{hmm:14} is formally described in Section~\ref{sec4}. To be self-contained, we include a pseudo-algorithm in the appendix. In Sections~\ref{sec5}-\ref{sec6}, numerical results are then presented to demonstrate the effectiveness of the reduced models. We close this paper with a short summary and discussion in Section~\ref{sec7}.

\section{Problem Statement and Background}\label{sec2}

We consider the nonlinear Schr\"odinger equation (NLS),
\begin{equation}
 i u_t = -u_{xx} + |u|^2 u, \label{nls}
\end{equation}
in one space dimension. For simplicity, we apply a periodic boundary conditions on a non-dimensionalized domain $x\in[0,\;2\pi]$. Here, the solutions of \eqref{nls} can be described by the Fourier series,
\begin{equation}
  u(x,t) = \sum_{k\in \mathbb{Z}} u_k(t) e^{ikx}.
\end{equation}
This turns the PDE into a set of ODEs for the Fourier modes,
\begin{equation}
 \frac{d}{dt}{u}_k = -i \omega_k u_k - i \sum_{k_1\in \mathbb{Z}} \sum_{k_2\in \mathbb{Z}} u_{k_1} u_{k_2} u_{k_1+k_2-k}^*, \label{ODE}
\end{equation}
with dispersion relation given by, $\omega_k=k^2.$ 

Of particular importance to the statistical mechanics interpretation of \eqref{nls} is the  Hamiltonian structure of the system, with the Hamiltonian given by,
\begin{equation}
 E= E_0 + E_1,\nonumber
\end{equation}
where,
\begin{equation}
\left\{
 \begin{aligned}
   E_0= & \sum_{k\in \mathbb{Z}} \omega_k |u_k|^2, \\
   E_1= & \frac{1}{2}\sum_{k_1 \in \mathbb{Z}} \sum_{k_2\in \mathbb{Z}} \sum_{k_3\in \mathbb{Z}} u_{k_1} u_{k_2} u_{k_3}^*u_{k_1+k_2-k_3}^*. 
 \end{aligned}
\right.\nonumber
\end{equation}
With this Hamiltonian, we can rewrite \eqref{ODE} as follows,
\begin{equation}\label{eq: uk}
i \frac{d}{dt}{u}_{k} =  \frac{\partial E}{\partial u_k^*}.
\end{equation}
Numerically, we can simulate the solutions of \eqref{eq: uk} with pseudo-spectral methods, e.g. \cite{bao2003numerical}, of \eqref{eq: uk} for finite wave numbers, $|k| \le K$. The initial condition can be prepared using a Monte-Carlo algorithm. We assume that the resulting solutions are the underlying dynamics.

In this paper, we are interested to construct a low-dimensional parametric model to predict low-frequency modes of \eqref{eq: uk}, given noisy observations of the corresponding modes at discrete-times. Namely, 
\begin{align}
v_{k,j} = u_k(t_j) + \varepsilon^o_j, \quad |k|\leq m,\label{obs} 
\end{align}  
where $m$ denotes the upper bound of the observed/resolved modes that are much smaller than the dimensionality of the underlying dynamics $K$, $m\ll K$. In \eqref{obs}, the noises $\varepsilon^o_j$ are i.i.d. Gaussian with mean zero and unknown error covariance, $R$. To achieve this goal, our strategy is to exhaust our physical knowledge of the model to deduce an appropriate ansatz for the parametric model and then apply a recently developed, adaptive ensemble Kalman filter based, parameter estimation method \cite{hmm:14} to specify the parameters in the corresponding model as well as the observation noise covariance, $R$. 

Before we discuss our main strategy, we briefly review a classical dimensional-reduction Mori-Zwanzig formalism \cite{Mori65,Zwanzig73,Zwanzig61}, which underpins the choice of ansatz for our parametric models in the remaining of this section. 

\subsection{Reduced Models from the Mori-Zwanzig formalism}\label{MoriZwanzig}

A general framework for reducing the dimension associated with a complex dynamical system is the Mori-Zwanzig projection formalism \cite{Mori65,Zwanzig73,Zwanzig61}, which was originally developed to deal with non-equilibrium processes in statistical mechanics. This approach relies on a projection operator, denoted by $\cal P$, which separates out the quantities of interest and identifies terms of different nature. In particular, for a system of initial value problem in the form, 
\begin{equation}
 \dot{x}= f(x), \quad x(0)=z,\label{ODE2}
\end{equation}
and an arbitrary reduced quantity, \(\varphi\), which is a function of $x(t)$, the Mori-Zwanzig procedure yields an exact equation for $\varphi$ \cite{Mori65,Zwanzig73}, 
\begin{equation}\label{eq: MZ}
    \frac{d}{dt} \varphi(t) = e^{t{\cal L}} {\cal{PL}} \varphi(0) + \int_0^t e^{(t-s){\cal L}} K(s) ds + \xi(t),
\end{equation}
where the first term in \eqref{eq: MZ} usually represents the reversible part of the dynamics and it represents the ``Markovian" term. Here, the differential operator $\cal{L}$ corresponds to the generator of the dynamical system in \eqref{ODE2} and it is defined with respect to initial condition $z$ as follows,  
\begin{equation}
\mathcal{L}= \sum_i  f_i (z)  \frac{\partial}{\partial z_i},
\end{equation}
and we use semigroup notation $e^{t\cal L}$ to denote the evolution operator that maps the solutions forward in time as follows, $\varphi(t) = e^{t\cal L}\varphi(0)$. The second term depends on $\varphi$ at all times between $0$ and $t$ so it incorporates the {\it memory effect} as a result of coarse-graining, and it dictates a strong coupling with the remaining degrees of freedom through a memory kernel,
\begin{equation}\label{eq: FK}
\quad K(t)={\cal{PL}}\xi(t),
\end{equation}
where 
\begin{equation}\label{eq: noise}
    \xi(t)= e^{t{\cal{QL}}} {\cal{QL}} \varphi(0), \quad \cal Q=I-\cal P.
\end{equation}
The term in \eqref{eq: noise} is referred to as the {\it orthogonal dynamics} and if the initial condition $z$ is random, then $\xi(t)$ is a stochastic forcing. Equation~\eqref{eq: MZ} is often called a \emph{generalized Langevin equation} (GLE). The most appealing aspect of the GLE in \eqref{eq: MZ} is that it is {\it exact}. However, solving the GLE in \eqref{eq: MZ} directly is not much simpler than solving the full system in \eqref{ODE2} since one has to estimate the orthogonal dynamics in \eqref{eq: noise} and the memory kernel function in \eqref{eq: FK}. 

We should point out that the GLE for the ODE in \eqref{ODE2} is nonunique since there are different choices for the projection operator $\cal P$. For example, in the work of Mori \cite{Mori65}, an orthogonal projection is employed, which is often appropriate when the problem can be formulated in a Hilbert space. For the NLS equation in \eqref{eq: uk}, Chorin~ et al \cite{chorin2000optimal} used a projection operator that is the conditional expectation with respect to the canonical ensemble $\rho \propto e^{-\beta E}$ to deduce an effective equation for selected (low-frequency) Fourier modes, $|k|\leq m$. This choice is motivated by the statistical mechanics aspect of the NLS \cite{lebowitz1988statistical}. Since the calculation is usually quite cumbersome, an expansion around the Gaussian distribution $\rho_0 \propto e^{-\beta E_0}$ was introduced, which is appropriate for systems at low temperature, $\beta \gg 1.$  To see this, one can introduce a change of variables in the Gibbs distribution, $ v= \sqrt{\beta} u.$ As a result, the distribution can be written as $\rho \propto e^{-E_0(v) - E_1(v)/\beta}.$ At low temperature when $\beta\gg 1,$ the distribution is approximately Gaussian. Furthermore, higher order terms in the equation are much less important, since  statistically, $u$ is of the order $1/\sqrt{\beta}.$

In the simplest case when $m=0,$ only the zeroth Fourier mode is retained and the  effective equation takes the form of \cite{chorin2000optimal},
\begin{equation}\label{eq: gle-u0}
\dot{u}_0 = - i c u_0 - i|u_0|^2 u_0 +  \int_0^t \kappa_0(t-\tau) u_0(\tau) d\tau
+ i \int_0^t \phi_0(t-\tau) |u_0(\tau)|^2 u_0(\tau) d\tau + \xi(t), 
\end{equation}
where $c$ is a positive constant,  and $\kappa_0$ and $\phi_0$ are complex valued kernel functions with complicated expressions \cite{chorin2000optimal} (they are written as infinite series). Furthermore, in solving \eqref{eq: gle-u0}, the history of the solution has to be stored and the integral has to be approximated by appropriate quadrature formulas {\it at each step} of the time integration. All these operations add up to significant computational cost and it is also unclear how the approximation by $\rho_0$ affects the modeling error. 

Rather than computing these kernels directly, we take a different approach here. In particular, we will model the memory terms in \eqref{eq: gle-u0} and the stochastic process $\xi$ with an appropriate ansatz of parametric equations. Subsequently, we estimate the corresponding parameters from noisy observations \eqref{obs} such that the resulting Markovian model gives accurate equilibrium statistical estimates for the selected Fourier modes that are retained: $|k|\leq m$.

\section{Constructing parametric models}\label{sec3}
Here we discuss our approach in approximating the GLEs using parametric models that involve explicitly few parameters.
These approximations are constructed in such a way that the approximate model can be re-written into a memory-less form, leading to a Markovian dynamics to facilitate the numerical implementation. To clarify the exposition, we first discuss the case where we only retain the zeroth mode. We assume the form of the GLE \eqref{eq: gle-u0}, but we approximate the memory terms using rational functions so that parameters can be introduced. We then provide the resulting parametric form for the more general case which retain more Fourier modes, $0<|k|\leq m$. 

\subsection{A reduced model for $u_0$ with scalar parametric approximation}

Now we will construct an ansatz for approximating the first memory term in \eqref{eq: gle-u0}. To this end, we introduce a parameter $b\in \bb{C}$ and an auxiliary function $f$ to denote the first memory term, 
\begin{equation}
b f :=   \int_0^t \kappa_0(t-\tau) u_0(\tau) d\tau,\label{bf}
\end{equation}
and our plan is to find a set of differential equations for solving $f$.  First, taking the Laplace transform on \eqref{bf}, we arrive at, 
\begin{equation}\label{eq: u2f}
  b \wt{f}(s)= \tilde \kappa_0(s) \wt{u}_0(s),
\end{equation}
where we denote $\widetilde{h}$ to be the Laplace transform (defined on frequency domain $s$) of any function $h$ that is locally integrable on $\mathbb{R}^+$.
The key idea is to approximate the kernel function, $\widetilde{\kappa}_0$, using a rational function,
\begin{equation}\label{eq: pade0}
\wt{\kappa}_0(s) \approx \frac{-|b|^2}{s-a},
\end{equation}
where $ a\in \bb{C}$ is the second parameter to be determined. This particular form of the rational function is chosen to ensure the stability of the resulting parametric model, as we will explain below. In principle, these two coefficients, $a$ and $b$, can be determined with Pad\'e approximations (or more general rational approximations) of the exact kernel. In model reduction problems, this is known as the {\it moment matching} procedure \cite{bai2002krylov,freund2003model}, where for linear dynamical systems, these parameters can be explicitly connected to properties of the original problem. The main departure of the current approach from those existing methods is mainly that we leave them as parameters and later infer them with a filtering procedure, learning from partially observed noisy time series. 

Converting \eqref{eq: u2f} and \eqref{eq: pade0} back to the time domain, we find that $f$ satisfies a differential equation,
\begin{equation}
\dot{f} = a f - b^* u_0(t), \quad f(0)=0.
\end{equation}
For low temperature case, one can neglect the second memory term in \eqref{eq: gle-u0} that involves $\psi_0$ since this higher-order term is negligible as explained before in Section~\ref{MoriZwanzig}. 
With this perspective, we propose the following parametric model,
\begin{equation}\label{eq: reduced-m0-I}
\left\{
\begin{aligned}
\dot{u}_0 = & - i c u_0 - i d |u_0|^2 u_0 + b f,\\
\dot{f} = & a f - b^* u_0 + \sigma_1 \dot{W}_f,
\end{aligned}
\right.
\end{equation}
where we have introduced two additional non-negative parameters $c$ and $d$. In principle, these parameters can be determined from the Mori-Zwanzig reduction procedure, which might involve lengthy calculations. However, since the derivation in \cite{chorin2000optimal} employed further approximations using the (conditional) Gaussian distribution, the resulting values for $c$ and $d$ may not be optimal. Therefore, we kept the form of the equations suggested by the Mori-Zwanzig formalism, but leave $c$ and $d$ as  additional parameters, which we will determine using a filtering procedure. 

We have also introduced a white noise $\dot{W}_f$. When the second equation is analytically solved and subsequently substituted into the first equation, this white noise will become a colored-noise approximation to the random process $\xi(t)$. When both $a$ and $b$ are real-valued parameters, the second equation represents an Ornstein-Uhlenbeck process \cite{uhlenbeck1930theory}. But here $f$ is a more general  Gaussian process. We should also point out that the reduced system of parametric equations in \eqref{eq: reduced-m0-I} is a special case of the physics constrained nonlinear regression model described in \cite{mh:13} in the following sense. In compact form, we can write \eqref{eq: reduced-m0-I} as a system of four-dimensional real valued SDEs,
\begin{align}
dx = [A x + N(x)] \,dt + \Sigma dW,\label{sde}
\end{align}
where we define $x = (\text{Re}\{u_0\}, \text{Im}\{u_0\},\text{Re}\{f\},\text{Im}\{f\})^\top$,  and $W$ is a standard two-dimensional Wiener process. In addition, we define $a=a_1 + i a_2$ and $b=b_1 + i b_2$ such that,
\begin{align}
A = \begin{pmatrix} 0 & c &  b_1 & -b_2  \\ 
 -c& 0 & b_2 & b_1  \\  
-b_1 & -b_2 & a_1 & -a_2  \\ 
b_2 & -b_1 & a_2 & a_1  \\ 
\end{pmatrix},\quad N(x)  = d\begin{pmatrix} |u_0|^2 \text{Im} u_0 \\
 -|u_0|^2 \text{Re}u_0 \\
  0 \\ 
  0 
\end{pmatrix}, \quad \Sigma = \begin{pmatrix} 0 & 0 \\ 0 & 0 \\ \frac{\sigma_1}{\sqrt{2}} & 0 \\ 0 & \frac{\sigma_1}{\sqrt{2}} \end{pmatrix}.
\end{align}

One of the main results in \cite{mh:13} states that if the Fokker-Planck operator of the SDE in \eqref{sde} is hypoelliptic, and suppose also that all eigenvalues of $A$ have negative real part and there exists an appropriate norm under which the inner product $\langle N(x),x \rangle=0$, then solutions of \eqref{sde} is geometrically ergodic. For our parametric model above, the stability condition is met when $a_1 \le 0$ and the dissipation of the energy of the nonlinear terms is satisfied under an inner product with respect to $L = E+\frac{1}{2}|f|^2$. 

We should note that one can repeat the  same calculation for approximating the second memory term in \eqref{eq: gle-u0} but the resulting nonlinear terms will not conserve energy and can be unstable (see Appendix A). Based on this consideration, we ignore approximating the second memory terms in this paper. Instead, we will only consider the parametric model in \eqref{eq: reduced-m0-I} which guarantees non-blowup solutions.  

\subsection{A reduced model for $u_0$ with multi-dimensional parametric approximations}

A simple extension of the two-parameter scalar parametric approximation model in \eqref{eq: reduced-m0-I} is to allow $b$ and $f$ to be vectors, here denoted by $\bm b$ and $\bm f$, respectively, to emphasize the multi-dimensional representation. This leads to an extended model,  
\begin{equation}\label{eq: reduced-m0-II}
\left\{
\begin{aligned}
\dot{u}_0 = & - i c \frac{1}{2} u_0 - i d |u_0|^2 u_0 + \bm b\cdot \bm f,\\
\dot{\bm f} = & A \bm f - \bm b^* u_0 + \Sigma \dot{W}.
\end{aligned}
\right.
\end{equation}
For instance,  the matrices $A$ and $\Sigma$ can  be chosen in the following form,
\begin{equation}
  A
  = 
  \left[
  \begin{array}{cc}
     a_1 & a_2\\
     -a_2 & a_1 
     \end{array}
     \right], \quad
  \Sigma
  = 
  \left[
  \begin{array}{cc}
     \sigma_1 & 0\\
     0 & \sigma_2
     \end{array}
     \right].
\end{equation}
The corresponding extended model has a parameter space of dimension 10. We will see that this model will give improved estimates compared to \eqref{eq: reduced-m0-I} in higher temperature case. Similar extensions can be found by increasing the dimension of $A$, $\bm b$ and
$\bm f$.  

\subsection{Models with more retained Fourier modes}
 In general, we can keep those modes $k$ with $|k|\leq m$, and $m$ indicates the range of the modes to be kept.
 In this case, we first define the coarse-grained energy,
 \begin{equation}
  E= \sum_{|k|\leq m} c_k |u_k|^2 + \frac{1}{2}\sum_{|k_1|\leq m} d_k \sum_{|k_2|\leq m} \sum_{|k_3|,|k_1+k_2-k_3|\leq m} 
   u_{k_1} u_{k_2} u_{k_3}^* u_{k_1+k_2-k_3}^*.   
\end{equation}
The model {\it without} memory can be written as follows,
\begin{equation}
  \dot{u}_k= -i \frac{\partial E}{\partial u_k^*}, \quad -m \le k \le m.
\end{equation}
Motivated by \eqref{eq: reduced-m0-I} and the Mori-Zwanzig procedure \cite{ChKaKu98}, let us consider a parametric model as follows,
\begin{equation}\label{multidimensional}
 \left\{
 \begin{aligned}
    \dot{u}_k=& -i \frac{\partial E}{\partial u_k^*} + b_k f_k, \\
     \dot{f}_k =& -b_k^* \frac{\partial E}{\partial u_k^*}  + a_k f_k + \sigma_k\dot{W}_k, \quad -m \le k \le m.
 \end{aligned}\right.
\end{equation}
The auxiliary functions are assumed to be zero initially, i.e., $f_k(0)=0,$ since they are introduced to approximate 
the memory terms.

To see the energy dissipation mechanism, we can define the Lyapunov functional,
\begin{equation}
  V = E + \sum_k |f_k|^2.
\end{equation}
Direct calculations yield,
\begin{equation}
 \frac{d}{dt}V = \sum_k \text{Re}(a_k) |f_k|^2.
\end{equation}
Consequently, we require that $\text{Re}(a_k)\le 0 $ to guarantee non-blow up solutions. 
 
\section{The Parameter Estimation Procedure}\label{sec4}
In this section, we describe formally how to estimate the parameters of the reduced models (e.g., \eqref{eq: reduced-m0-I}, \eqref{eq: reduced-m0-II}, or \eqref{multidimensional}),
given noisy observations $v_j = (v_{-m,j},\ldots,v_{m,j})^\top\in\mathbb{C}^{2m+1}$ of $u_j=(u_{-m,j},\ldots,u_{m,j})^\top\in\mathbb{C}^{2m+1}$, where $u_{k,j}=u_k(t_j)$ are solutions of the full system in \eqref{ODE} for $|k|\leq K$ and $K\gg m$ at discrete time step $t_j$:
\begin{align}
v_{j} = u_{j} + \epsilon_{j}, \quad \epsilon_{j}\sim\mathcal{N}(0,R), \quad |k|\leq m,\label{noisyobs}
\end{align}
with an unknown observation error covariance $R$, where $R$ is an $(2m+1)\times (2m+1)$ diagonal matrix with $k$-th diagonal component, $r_k$. To simplify the discussion, let us classify the parameters in our reduced model to two types. We refer to the parameters in the deterministic terms in the reduced models as the ``deterministic parameters", $\theta_d$, and the amplitude of the stochastic forcings as the ``stochastic parameters", $\theta_s$. For example, in \eqref{eq: reduced-m0-I}, the deterministic and stochastic parameters are, $\theta_d=\{a, b,c,d\}$, and $\theta_s=\{\sigma_1^2,R\}$, respectively. We split the parameters into two types because the algorithm that we use to estimate $\theta_d$ is simply a standard augmentation method while the algorithm to estimate $\theta_s$ is the adaptive method for estimating covariances which preserves the positivity of $\sigma_1^2$ and $R$. Let us also define vector $x_j = (u_{-m,j},\ldots,u_{m,j}, f_{-m,j},\ldots,f_{m,j})^{\top}\in\mathbb{C}^{4m+2}$ to simplify the notation below.

The main idea of the parameterization method is to apply Bayes' theorem to obtain a posterior distribution of the augmented state and parameters at each time step $t_j$ when observations become available,  
\begin{align}
p(x_j,\theta_d,\theta_s| v_j)&\propto p(x_j,\theta_d,\theta_s) p(v_j| x_j,\theta_d,\theta_s),\label{bayes}
\end{align}
where $p(x_j,\theta_d,\theta_s)$ denotes the prior distribution of the augmented state and parameters at time $t_j$ and $p(v_j| x_j,\theta_d,\theta_s)$ denotes the likelihood function of the augmented state and parameters, corresponding to the observation model in \eqref{noisyobs}, that is, 
$p(v_j| x_j,\theta_d,\theta_s) = \mathcal{N}(u_j,R)$. The parameterization method can be formally described as follows: Since $p(x_j,\theta_d,\theta_s) = p(\theta_s)p(x_j,\theta_d|\theta_s)$ by definition of the conditional distribution, we can rewrite \eqref{bayes} as follows:
\begin{align}
p(x_j,\theta_d,\theta_s| v_j)&\propto  p(\theta_s) p(x_j,\theta_d|\theta_s) p(v_j| x_j,\theta_d,\theta_s),\label{bayes2}\\
&\propto p(\theta_s) p(x_j,\theta_d|v_j,\theta_s),\label{bayes3}
\end{align}
where we use another Bayes' theorem, $p(x_j,\theta_d|v_j,\theta_s)\propto p(x_j,\theta_d|\theta_s) p(v_j| x_j,\theta_d,\theta_s)$, to obtain \eqref{bayes3}. Here, the first step in the filtering algorithm is to estimate $p(x_j,\theta_d|\theta_s,y_j)$ by applying Bayes' theorem to the last two components of \eqref{bayes2}. Subsequently, we implement the Bayes' theorem one more time in \eqref{bayes3} to obtain the posterior distribution of the augmented $(x_j,\theta_d,\theta_s)$.


To avoid unobservability of the stochastic parameters due to sparse observations with dimension less than the number of stochastic parameters, $\theta_s$, in our implementation, we include information from past observations up to lag $L>1$.  At each time step $t_j$, instead of solving \eqref{bayes2}-\eqref{bayes3}, we formally solve
\begin{align}
p(x_j,\theta_d,\theta_s| v_j,\ldots,v_{j-L+1}) &\propto p(x_j,\theta_d,\theta_s|v_{j-1},\ldots,v_{j-L+1}) p(v_j| x_j,\theta_d,\theta_s) \nonumber \\ &\propto p(\theta_s) p(x_j,\theta_d|\theta_s,v_{j-1},\ldots,v_{j-L+1}) p(v_j| x_j,\theta_d,\theta_s)\label{bayes4}\\
&\propto p(\theta_s) p(x_j,\theta_d|\theta_s,v_j,\ldots,v_{j-L+1}),\label{bayes5}
\end{align}
where, similar as before, the first step is to estimate $p(x_j,\theta_d|\theta_s,v_j,\ldots,v_{j-L+1})$ by applying Bayes' theorem to the last two components of \eqref{bayes4}. Subsequently, we implement the Bayes' theorem one more time in \eqref{bayes5} to obtain the posterior distribution of the augmented $(x_j,\theta_d,\theta_s)$.

In our numerical implementation, we use the method in \cite{hmm:14} which uses Gaussian approximation to solve these inverse problems. At time $j\geq L$, we assume that we have prior ensemble estimates of $\{x_j,\theta_d,\theta_s\}$ at times $j, j-1, \ldots, j-L+1$ and the associated observations at these times. We assume that the deterministic parameter is persistence, that is, $\dot \theta_d= 0$. The first step is to apply ETKF method \cite{hunt:07} to obtain posterior ensemble estimates of the augmented variable $\{x_j,\theta_d\}$, incorporating observations in \eqref{noisyobs}, which is a Gaussian approximation of $p(x_j,\theta_d|\theta_s,v_j,\ldots,y_{j-L+1})$. To start the algorithm, one can just repeat this ETKF algorithm $L$-times to obtain the prior ensemble estimates at time $j=1,\ldots,L$ with fixed parameters $\{\theta_s,\theta_d\}$. Now, at $j\geq L$, we start the secondary filter to update $\theta_s$.
The key idea of the secondary filter is to view the posterior density function $p(x_j,\theta_d|\theta_s,v_j,\ldots,v_{j-L+1})$ as a likelihood function of $\theta_s$. Notice that while this posterior density is Gaussian with respect to variables $(x_j,\theta_d)$, 
its dependence on $\theta_s$ can be described  non-uniquely (for example, see \cite{mehra:70,mehra:72,belanger:74,bs:13,hmm:14,zh:14}). Here, we will adopt the estimation method of \cite{hmm:14} that is based on Belanger's formulation \cite{belanger:74} with a likelihood function corresponding to the following pseudo-observation model,
\begin{align}
\sigma_{j,\ell} =  \mathcal{F}_{j,\ell}\theta_s + \eta_{j,\ell},\quad \eta_{j,\ell}\sim\mathcal{N}(0,W_{j,\ell}), \quad \ell =j,\ldots, j-L+1\label{pobsmodel}
\end{align}  
Here, components of $\sigma_{j,\ell} = \{e_j e_{j-\ell}^\top\}$ are the product of the forecast error estimates in the observation space (which are also known as innovations),
\begin{align}
e_j = v_j - \bar{u}^-_j,
\end{align}
where $\bar{u}^-_j$ denotes the mean prior estimate that is empirically estimated with an ensemble average. In \eqref{pobsmodel}, the observation operator $\mathcal{F}_{j,\ell}$ and the noise covariance matrix $W_{j,\ell}$ are functions of $\bar{x}^-_{j-\ell}$ and $\theta_d$ and they will be constructed recursively. We should also note that in our implementation, $W_{j,\ell}$ is approximated under a Gaussian assumption (see Appendix B below for detail). With the pseudo-observation model in \eqref{pobsmodel}, a secondary Kalman filter is implemented $L$-times to sequentially update the posterior mean and covariance estimate of $\theta_s$, accounting for pseudo-observations $\{\sigma_{j,\ell}\}_{\ell=1\ldots, L}$ one at a time. To be self-contained, we provide a pseudo algorithm of this method in Appendix B below. We should note that there are other methods to approximate the secondary Bayes' update in \eqref{bayes5} that use different observation model  in \eqref{pobsmodel} and do not use Kalman update (see e.g., \cite{bs:13,zh:14}). 

\section{Numerical results on parameter estimation of  models with a single Fourier mode $u_0$}\label{sec5}
In this section, we present the results from three numerical tests, where the parametric models for $u_0$ (equations  \eqref{eq: reduced-m0-I}  and \eqref{eq: reduced-m0-II}) are estimated and further assessed.   We assume that the observation time interval $T_{obs}=0.02$. The time series, consisting of $100,000$ observations, is generated by using the Strang's splitting method in time, which has been implemented in \cite{bao2003numerical} for the NLS equation.   The data is generated from the solutions of the NLS in the Fourier domain \eqref{ODE}, with $K=32$.  

We then perform a parameter estimation method with an ensemble Kalman filter based method \cite{hmm:14}, which was described in the previous section.  The forecast is generated with the 4th order Runge-Kutta method with step size $\Delta t=0.002.$ 

Following the estimation, we verify the forecasting skill of the reduced models with the estimated parameter set as follows. We take the estimated parameters and run the reduced models forward in time for a sufficiently long time. Then, based on the long trajectory, we compare the low-order statistics up to order-four and the time correlation function. The auto-correlation is  computed based on the Wiener-Khinchin theorem. Namely, we take the Fourier transform of the data, $x(t),$ and compute the power spectrum, $|\hat{x}(\omega)|^2.$ The correlation function is then given by  the inverse Fourier transform of the power spectrum. This procedure is often more efficient than the direct approach, i.e.,
$$c(\tau) \approx \frac{1}{M} \sum_{m=1}^M x(t_m+\tau) x(t_m) .$$

\subsection{Low temperature $\beta=10^{4}.$}

\begin{figure}
\centering
\includegraphics[scale=0.5]{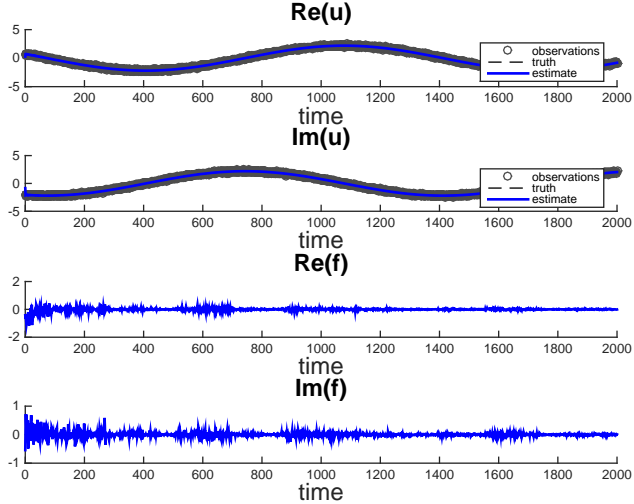}
\caption{State estimates $u$ and $f$ for the parametric model \eqref{eq: reduced-m0-I}.}
\label{param_fig1}
\end{figure}

\begin{figure}
\centering
\includegraphics[width=0.6\textwidth]{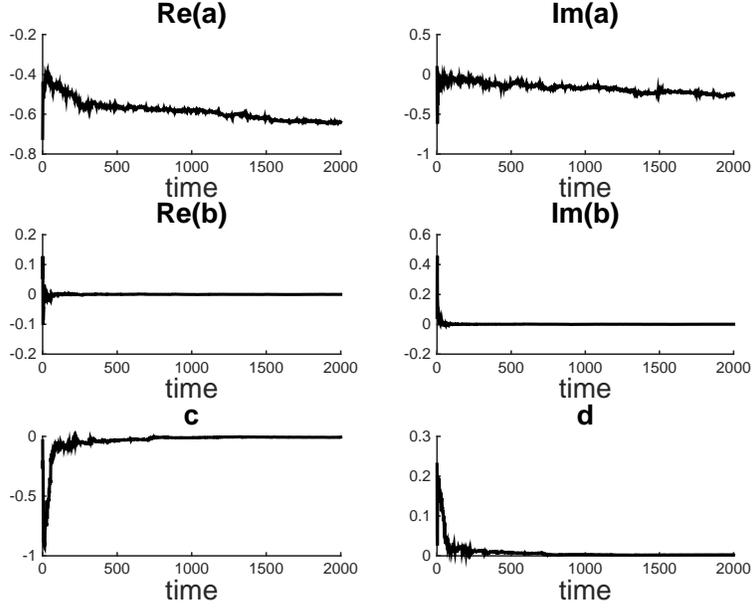}
\caption{Deterministic parameter estimates, $a$, $b$, $c$ and $d$.}
\label{param_fig2}
\end{figure}

\begin{figure}
\centering
\includegraphics[scale=0.5]{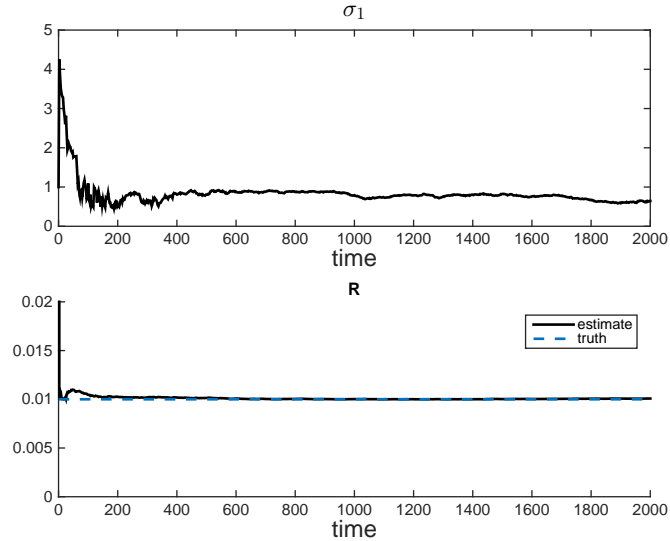}
\caption{Stochastic parameter estimates, $\sigma_1, {R}$.}
\label{param_fig3}
\end{figure}

We first consider a low temperature case with $\beta= 10^{4}$ ($k_B T=0.0001$), and estimate the parameters in the model \eqref{eq: reduced-m0-I} for $u_0$. This model contains two (complex) deterministic parameters $a$ and $b$, together with two real valued parameters $c$ and $d$. We set the observation noise error with variance, ${R}=0.01$.  In Fig. \ref{param_fig1}, we show the estimated solutions along with the observed values during the estimation procedure. Very good agreement with the true values has been found. We also monitor the predicted values of the deterministic parameters, and the history is presented in Fig. \ref{param_fig2}. Meanwhile, the stochastic parameter $\sigma_1$ has settled to a constant value, and the variance of the observation noise  has been correctly predicted, as indicated in Fig. \ref{param_fig3}. Another observation is that the predicted values of the parameters $c$ and $d$ 
are $c=-0.0067$ and $d=0.0024$, which are quite different from the values calculated from the Mori-Zwanzig's projection procedure ($0.00063$ and $1$, respectively) corresponding to an approximate Gaussian measure, $e^{-\beta E_0}$. We should point out that if we fix these two parameters to be those from the Mori-Zwanzig projection and run the filtering procedure to estimate the remaining parameters, $a, b, \sigma_1, R$, the resulting estimates are completely innaccurate. This suggests that while the perturbation approach \cite{chorin2000optimal} suggests the explicit forms of the reduced model, it is more natural to adaptively estimate all the parameters which reconfirms the results in \cite{bh:14}. Moreover, in general, there can be non-unique parameters that provide the same equilibrium statistics (for e.g., see Proposition 1(d) in \cite{hmm:14}). 


\begin{figure}
\centering
\includegraphics[width=0.8\textwidth]{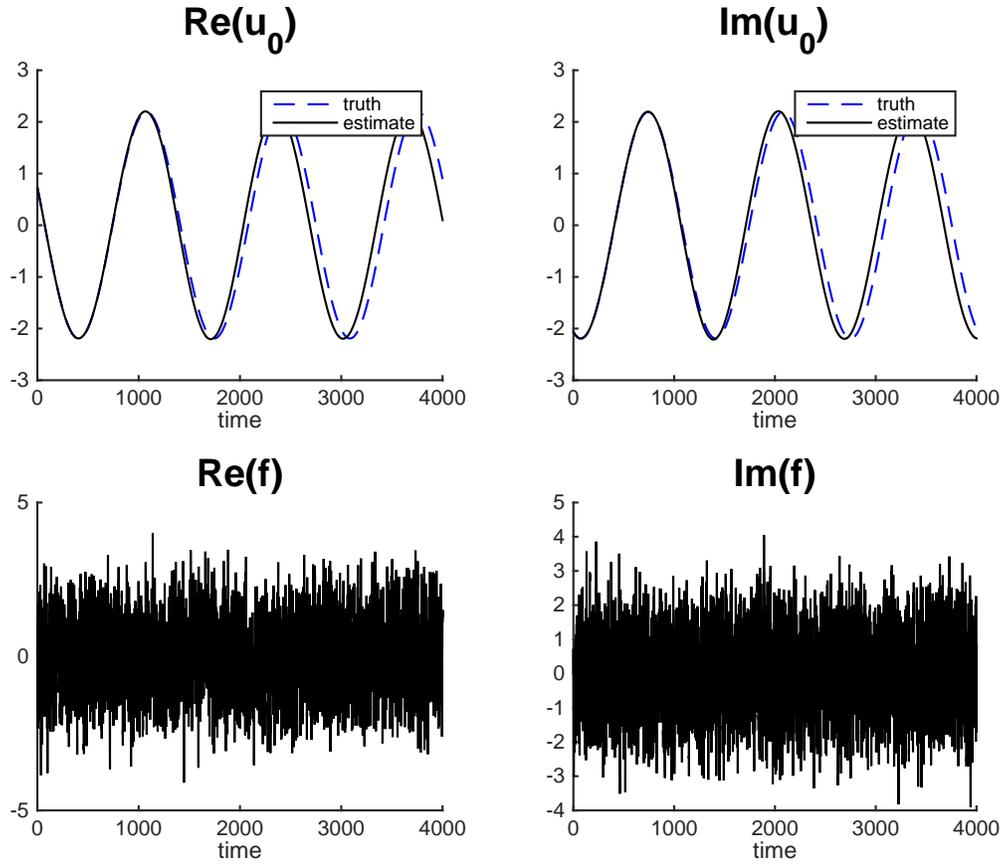}
\caption{Solutions of the reduced model in \eqref{eq: reduced-m0-I}, integrated with the estimated parameters.}
\label{verify1}
\end{figure}
\begin{figure}
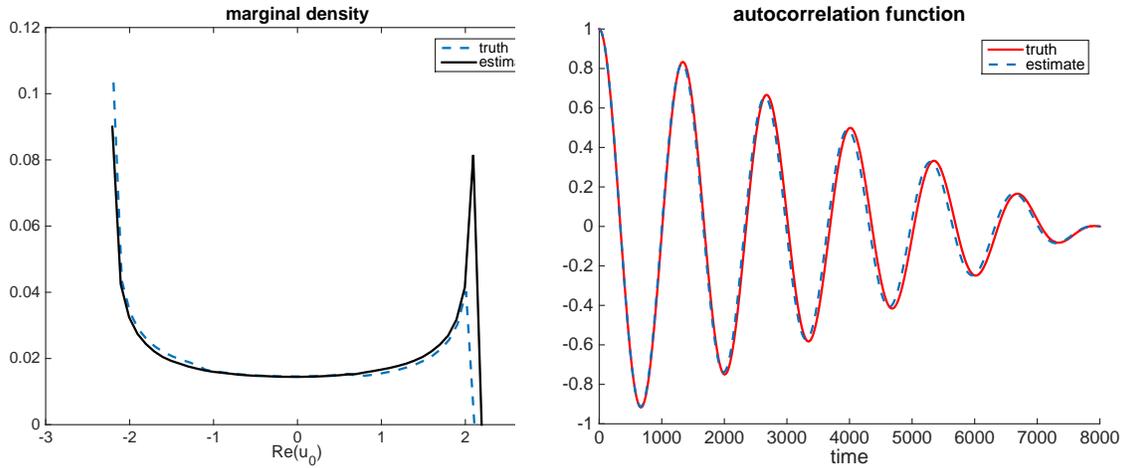

\centering
\includegraphics[width=.44\textwidth]{verification_fig2b.eps}
\includegraphics[width=.45\textwidth]{verification_fig3b.eps}
\caption{The marginal distribution (left) and the time correlation function (right) predicted by the reduced mode \eqref{eq: reduced-m0-I} for low temperature, $\beta=10^4$. As comparison, the statistics of the true solution is also shown.}
\label{verify2}
\end{figure}

Next we evaluate the forecasting skill and check the accuracy of the climatological statistics of the resulting solution $u_0$. We observe from Fig. \ref{verify1} that the qualitative behavior of the path-wise solutions is well captured. The solution $f$, which was introduced to replace the memory term, exhibits much faster oscillations.   Further, from Fig. \ref{verify2}, we observe that the distribution and the time correlation of the true solution are  accurately predicted. We also report the accuracy of the first four moment estimates in Table~\ref{tab0}, where the errors are on the order of $10^{-2}$. 


\subsection{Results for a higher temperature $\beta=10.$}

We now turn to observations obtained from a higher temperature simulation with $\beta=10$ ($k_BT=0.1$). In this case, the time series for $u_0$ exhibits faster oscillations and slightly larger amplitude. We also observe the amplitude and frequency of the oscillations are somewhat sensitive to the initial conditions since the system is not ergodic.

\begin{figure}
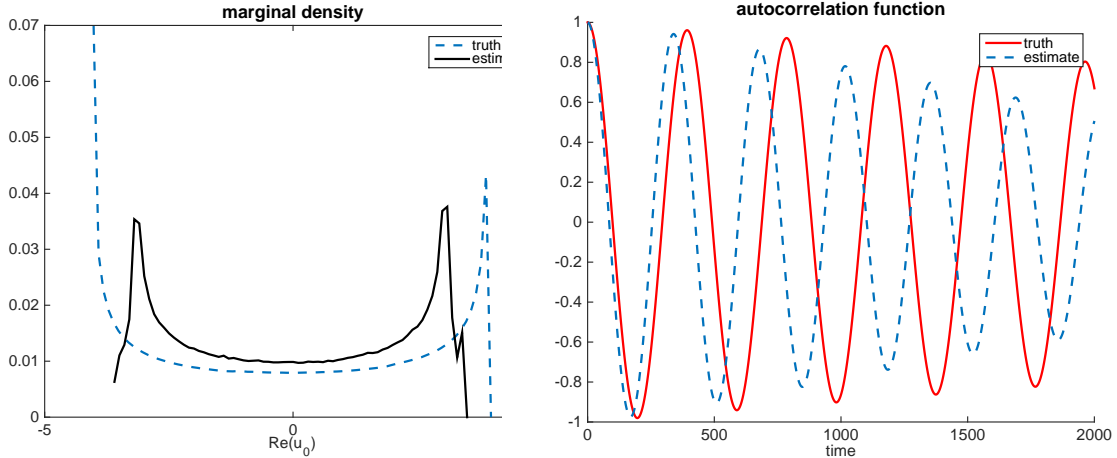

\centering
\includegraphics[width=.434\textwidth]{verification_fig2_kt01.eps}
\includegraphics[width=.45\textwidth]{verification_fig3_kt01.eps}
\caption{Predicted marginal distribution and correlation function for $\beta=10$ using the first reduced model \eqref{eq: reduced-m0-I}.}
\label{verify01-rk4}
\end{figure}

Based on the data, we estimate the first  reduced model \eqref{eq: reduced-m0-I}, and then check the statistics of the resulting model. 
Due to the higher temperature, the variance of $u_0$ is much bigger. Therefore, we set a higher value for observation noise variance, ${R}=0.1$. We see from Fig. \ref{verify01-rk4} that  the accuracy is not as satisfactory as in the previous case. In particular, the peaks of the marginal density are not well captured and the variance is underestimated (see Table~\ref{tab0}) although the other statistics are accurately estimated. Also, the correlation function is inaccurate beyond the first oscillation.

\begin{figure}
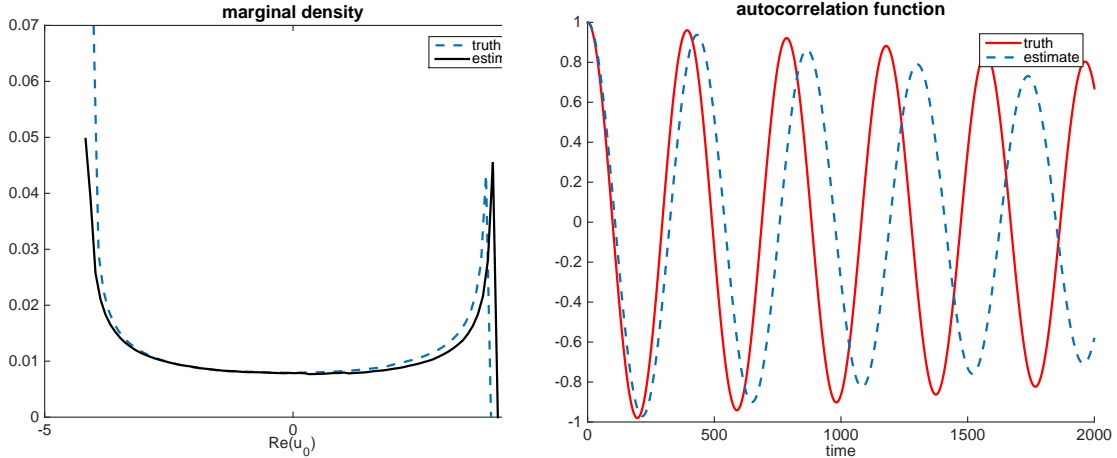

\centering
\includegraphics[width=.434\textwidth]{verification_fig2_rk6d.eps}
\includegraphics[width=.45\textwidth]{verification_fig3_rk6d.eps}
\caption{Predicted marginal distribution and correlation function for $\beta=10$ using the reduced model \eqref{eq: reduced-m0-II} .}
\label{verify01-rk6d}
\end{figure}

As comparison, we consider the next parametric model, represented by the equation \eqref{eq: reduced-m0-II}, which  contains 4 complex deterministic parameters and two real ones. With the parameters obtained from the filtering procedure, we perform a similar statistical
verification. The results, including the histogram and the time correlation functions,  are illustrated in Fig. \ref{verify01-rk6d}. It is clear that the extension has offered improved accuracy in the resulting histogram.  Table \ref{tab0} summarized
the statistics (four moments) of $u_0$ obtained from the three tests, compared to the true values. We notice that for the higher temperature case, the model \eqref{eq: reduced-m0-II} yields much better estimates for the second moment. However, the estimated correlation function is only slightly improved up to time $500$ relative to the result from the model in \eqref{eq: reduced-m0-I}.

\begin{table}[htdp]
\caption{Comparison of the equilibrium statistics  of $\text{Re}(u_0)$ for the three tests.}
\begin{center}
\begin{tabular}{|c|c|c|c|c|c|c|}
\hline
    & \multicolumn{2}{|c}{Model \eqref{eq: reduced-m0-I}, $\beta=10^4$} & \multicolumn{2}{|c}{Model \eqref{eq: reduced-m0-I}, $\beta=10$}& \multicolumn{2}{|c|}{Model \eqref{eq: reduced-m0-II}, $\beta=10$}\\
\hline
    Statistics &    Truth & Estimate &  Truth &  Estimate & Truth& Estimate \\
    \hline
    mean     & -0.0037  &  -0.0717    & 0.0457 &   0.0368 &   0.0457  & -0.0642\\
    \hline
      variance     & 2.4018   &2.4570     & 8.0655  &  5.3181&    8.0655  & 8.5148 \\
    \hline
    skewness     &0.0840  & 0.0658   & -0.0251  &  -0.0204 &   -0.0251 &  0.0371 \\
    \hline
    kurtosis     & 1.5071 &   1.5123  &  1.4998 &   1.5160 &  1.4998  &  1.5062\\
    \hline
   \end{tabular}
\end{center}
\label{tab0}
\end{table}

\bigskip

\section{Numerical results for multiple retained Fourier modes}\label{sec6}
\begin{figure}
\centering
\includegraphics[width=.9\textwidth]{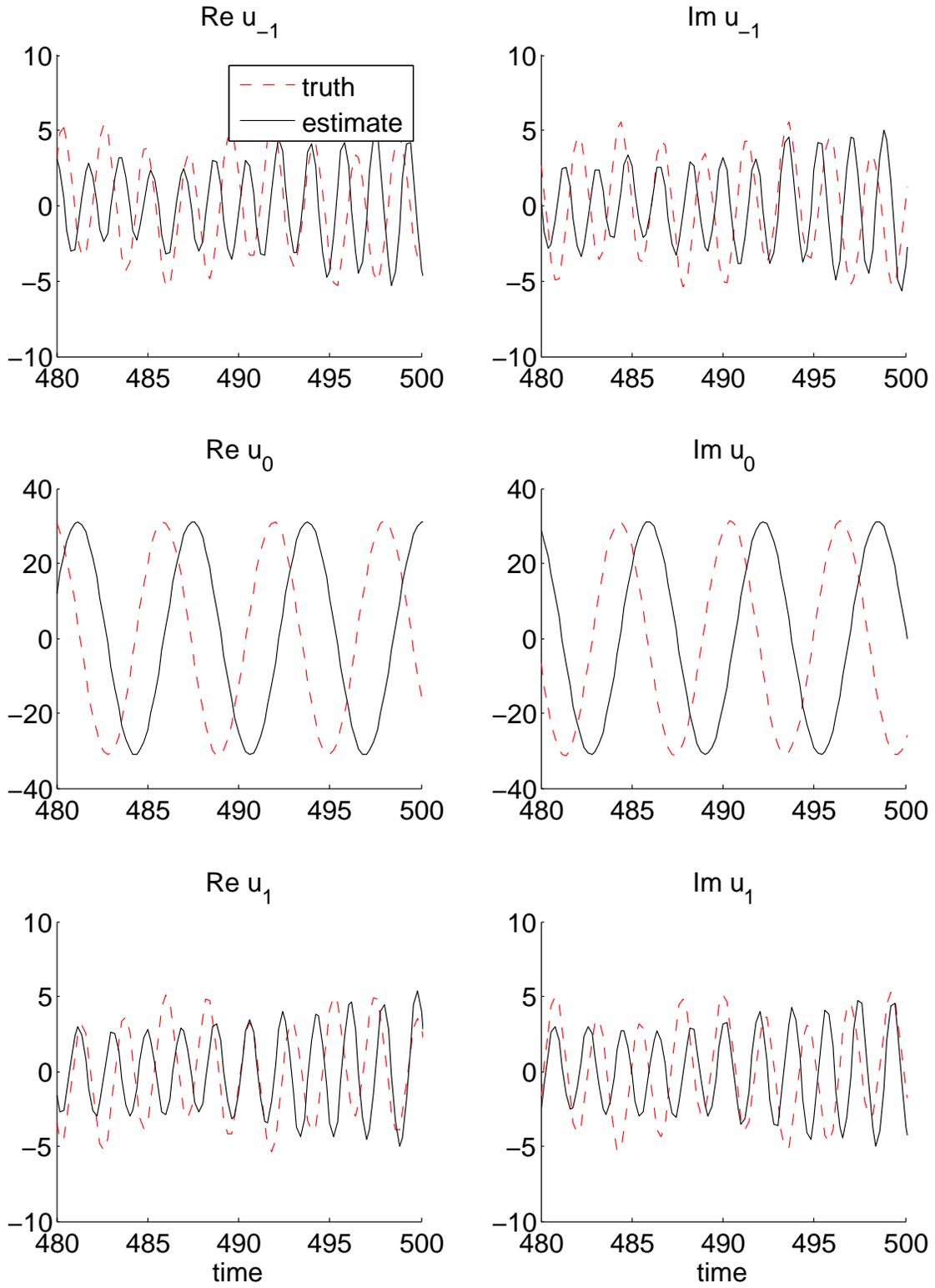}
\caption{Solutions of the reduced model in \eqref{multidimensional}, compared to those of the full model.}
\label{verify00-rk12d}
\end{figure}

In this section, we consider modeling three modes, $u_{-1},$ $u_0,$ and $u_1$, in the Fourier series for much higher temperature case with $\beta=1/20$ ($k_BT=20$). In this numerical experiment, the parametric model in \eqref{multidimensional} has 6 dimensional complex valued variables and 18 real valued parameters. For this case, we found several numerical issues when including more Fourier models in the reduced models. First, Fourier modes $u_{-1}$ and $u_1$ exhibit very different frequency compared with that of $u_0$, which can be seen in Fig. \ref{verify00-rk12d}. As a consequence, the variance of each component has different scales (see Table~\ref{table2}). The disparity in covariance scaling becomes exceedingly large when the temperature is low. As a result, much smaller time steps are needed in the estimation procedure to sample the observations for $u_{-1}$ and $u_1$. On the other hand, the procedure has to be continued for a long time period to make sufficient observations of $u_0$. A more flexible estimation method would be more useful in this case. 

A second related issue is that these three multiscale Fourier modes are correlated and this suggests that one may need a different ansatz for the parametric models. For example, one may need to consider fully correlated noises in the equations for $f_k$ in \eqref{multidimensional}, which means more parameters to fit. An alternative way to overcome this issue is to fit the model in \eqref{multidimensional} to the uncorrelated observations that can be obtained by rescaling the observations with the covariance matrix. In particular,
we define our observations as follows,
\begin{align}
v_j = \tilde{u}_j + \varepsilon_j, \quad\varepsilon\sim\mathcal{N}(0,R),\label{rescaledobs}
\end{align}
where $\tilde{u}_j \equiv C^{-1/2}u_j$ is the rescaled of $u_j$ by the equilibrium covariance matrix $C$ that can be computed empirically through time averaging of a long time series, assuming the stationarity and ergodicity of the underlying dynamics. Note that this rescaling improves the identifiability of $u_0$ that has much larger variance relative to $u_1, u_{-1}$ (again, see Table~\ref{table2}) since the equilibrium covariance of the rescaled variables $\tilde{u}_j$ is an identity covariance matrix, $\mathcal{I}$. In our numerical experiment below, we assume that the observation error covariance to be 10\% of identity, $R=0.1\mathcal{I}$ and we fit the rescaled observations in \eqref{rescaledobs} to \eqref{multidimensional}, where the observation time interval is chosen to be $T_{obs}=0.02$ and the training data set is 100000 data points. 

To confirm the success of the filtering procedure, we see that the filter estimate for $R$ converges to the true value, $R=0.1\mathcal{I}$, and the filter estimates for $\tilde{u}_j$ have an equilibrium covariance that is indeed identity, exactly equals to the equilibrium covariance of $\tilde{u}_j$. The last point here, however, does not imply that the solutions of the reduced model in \eqref{multidimensional}, integrated with the estimated parameters, will have an identity equilibrium covariance matrix. To verify the predictive skill of the resulting parametric model in \eqref{multidimensional}, we rescale the solutions, $\breve{u}_j$, to the appropriate scaling of the underlying dynamics with the following covariance transformation, $\hat{u}_j=C^{1/2}\tilde{C}^{-1/2}\breve{u}_j$, where $\tilde{C}$ is the equilibrium covariance of $\breve{u}_j$. In our numerical experiment, we compute this statistics, $\tilde{C}$, by averaging the solutions, $\breve{u}_j$, of \eqref{multidimensional} at 6000-10000 model time units, at discrete time step of $T_{obs}=0.02$. In particular, the solutions $\breve{u}_j$ are obtained by integrating the model in \eqref{multidimensional} with parameters determined by averaging over the last 1000 steps of the filter estimates.

\begin{figure}
\centering
\includegraphics[width=.7\textwidth]{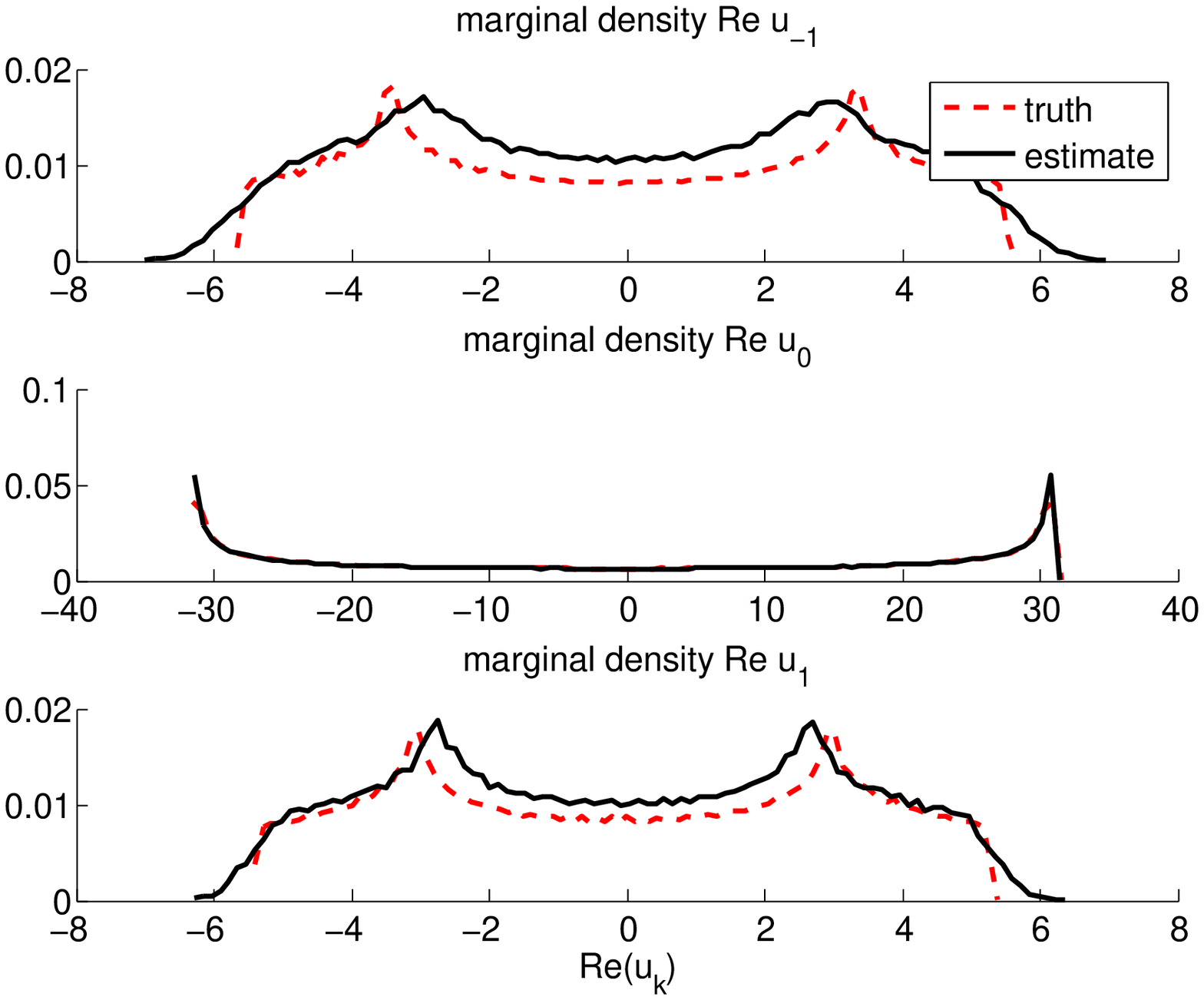}
\includegraphics[width=.7\textwidth]{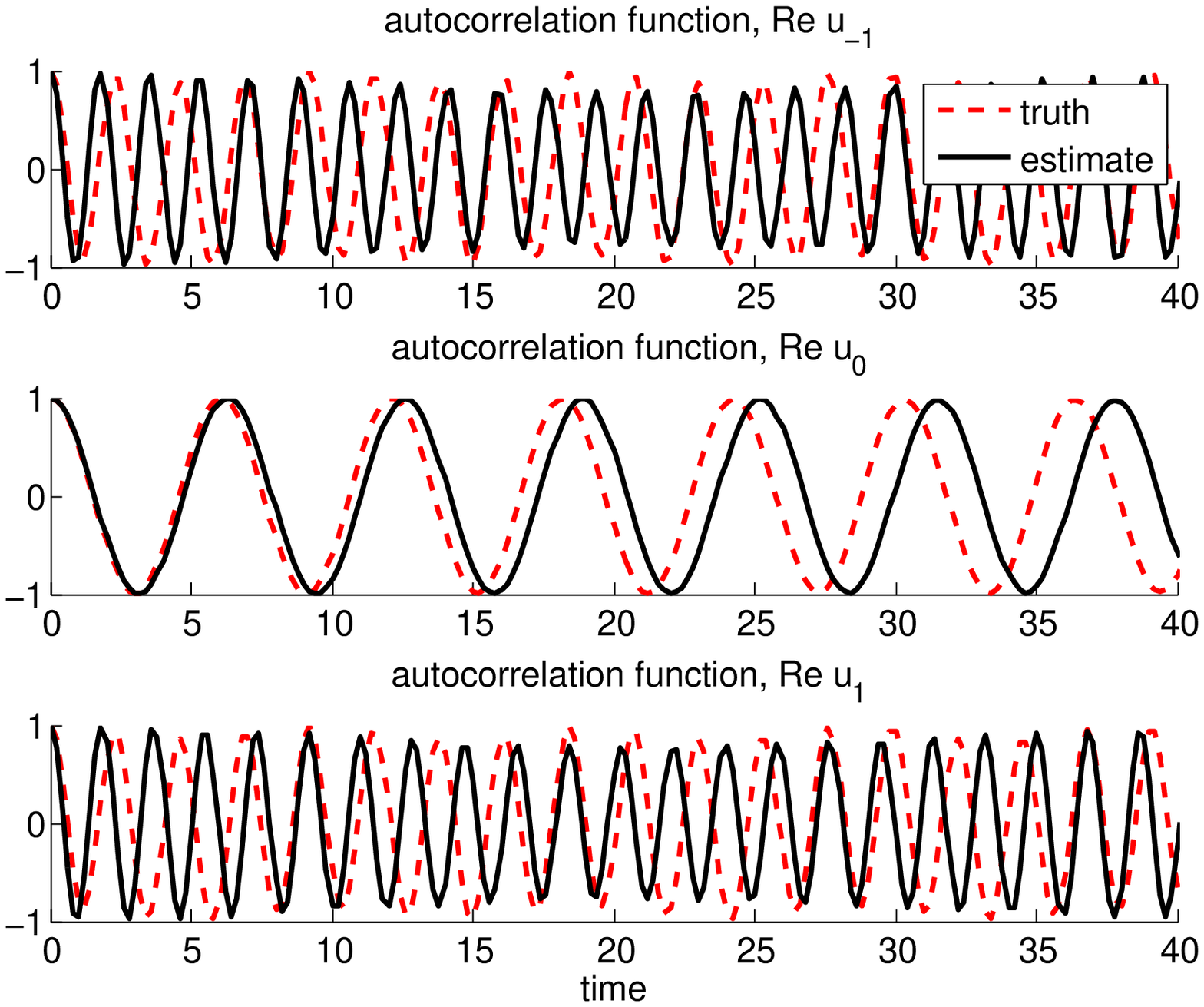}
\caption{Predicted marginal distributions and correlation functions for $\beta=1/20$ from the reduced model in
\eqref{multidimensional}}
\label{verify01-rk12d}
\end{figure}

In Fig. \ref{verify00-rk12d}, an example of the solutions from the parametric model \eqref{multidimensional} is shown; here we compare the estimates $\breve{u}_j$ (black solid line) with the truth $u_j$ (red dashes) at an arbitratry period of time interval. Notice that even if we don't expect a path-wise agreement, the qualitative behavior of the solutions are reasonably reproduced (in the sense that their magnitude and frequency are qualitatively comparable). In Fig.~\ref{verify01-rk12d} the comparison of the histogram and the time correlation functions to those of the full model is demonstrated. Notice that despite the difference in amplitude  and temporal scalings between modes $u_0$ and $u_{-1}, u_1$, the nontrivial marginal distributions are well captured. The correlation times for mode $u_0$ are well captured at least until 10 unit time; for the other modes, $\{u_{-1},u_1\}$, the correlation times are in agreement for about one period of oscillation (approximately up to one unit time). We also report the first four moments estimates compared to those of the truth for each variables in Table~\ref{table2}. The exact agreement in terms of variances are not surprising since we purposely scale the estimates to match the covariance of the true dynamics. However, the agreement in terms of the higher order moments such as skewness and kurtosis is nontrivial.


\begin{table}[htdp]
\caption{Equilibrium statistics predicted by the model in \eqref{multidimensional} compared to those of the full model.}
\begin{center}
\begin{tabular}{|c|c|c|c|c|c|c|}
\hline
& \multicolumn{2}{|c|}{Re$u_{-1}$} & \multicolumn{2}{|c|}{Re$u_{0}$} & \multicolumn{2}{|c|}{Re$u_{1}$} \\
\hline
statistics & truth & estimate & truth & estimate & truth & estimate \\ \hline
mean & 0.0008 & 0.0005 & -0.0036 & 0.0164 & 0.0019 & -0.0004 \\
variance & 10.5646 & 10.5646 & 487.4128 & 487.4128 & 9.3998 & 9.3998 \\
skewness & -0.0002 & -0.0003 & 0.0004 & -0.0009 & -0.0007 & -0.0002 \\
kurtosis & 1.6594 & 1.8106 & 1.5000 & 1.5000 & 1.6999 & 1.8017\\
\hline
\end{tabular}
\end{center}
\label{table2}
\end{table}%

\section{Summary and Discussion}\label{sec7}
This paper presented a modeling approach that blends some physical knowledge about the underlying dynamics and the availability of training data to predict low-frequency modes of the NLS equation. In particular, we use the Mori-Zwanzig formalism as guidelines to construct effective parametric models and apply an adaptive ensemble Kalman filter to estimate the parameters. The novelty here is that we approximate the memory term and the orthogonalized dynamics of a generalized Langevin equation obtained from the Mori-Zwanzig expansion with a rational function and a colored noise, respectively. It turns out that the resulting parametric model here is an example of the physics constrained nonlinear regression modeling approaches proposed in \cite{mh:13,hmm:14}. This serendipity allows one to use the stability conditions established in \cite{mh:13} to ensure non-blow up solutions of the resulting parametric model.  Compared to the full GLE, these models have advantages  in practical implementations because they do not involve memory.   

The climatological forecasting skill of the proposed parametric model was verified in terms of the first four moments, marginal densities, and correlation functions for various temperatures. For low temperature case, high predictive skill of Fourier mode $u_0$ is obtained with a reduced model with a scalar parameterization for the memory term \eqref{eq: reduced-m0-I}. For higher temperature case where the scale-gap is smaller than the low temperature case, the problem becomes more challenging. In this situation, we showed that one can improve the estimates either with a two-dimensional parameterization for the memory term in \eqref{eq: reduced-m0-II} or with fitting more modes into a model with  more retained modes in \eqref{multidimensional}. 

With the encouraging results in this paper, we plan to apply this modeling strategy on other applications such as on coarse-grained biomolecular models \cite{IzVo06,lange2006collective,oliva2000generalized,LiXian2014} in our future research. 
In general problems, however, the success of this modeling approach will depend mostly on the choice of the ansatz for modeling the memory terms. As it has been theoretically established in \cite{bh:14}, if the ansatz is adequate, then it is possible to obtain, both, accurate climatological statistical forecasting and optimal filtering. Our NLS example in this paper empirically suggested that our ansatz is optimal in this case. Other potential issue is in the parameter estimation strategy which can be expensive when more observations are included. While many cheaper parameterization methods are available (such as regression-based or maximum likelihood-based algorithms), these methods are often inferior to the adaptive method applied in the present work even when adequate ansatz is used as shown in \cite{bh:14}. Therefore, improving the numerical efficiency of the adaptive parameter estimation scheme that we used here \cite{hmm:14} or its variant (see e.g., \cite{bs:13,zh:14}) will be the key for successful applications in more complex problems.

\begin{acknowledgments}
The research of JH is partially supported by the the ONR MURI grant N00014-12-1-0912, ONR grant N00014-13-1-0797, and the NSF grant DMS-1317919.
\end{acknowledgments}


%

\appendix

\section{Remarks on the second memory terms in \eqref{eq: gle-u0}}
Mathematically, one can also include the second memory term using the similar rational approximation for the high temperature case when this term is not negligible. Denoting the other kernel function as,
\begin{equation}
\beta g(t) =  i \int_0^t \phi_0(t-\tau) u_0(\tau) |u_0(\tau)|^2 d\tau, 
\end{equation}
where $\beta$ is an additional parameter and approximate the Laplace transform of the kernel function,
\begin{equation}
\wt{\phi}_0(s) \approx r_\phi(s)= \frac{-\beta^2}{s-\alpha},
\end{equation}
where we assume that $\alpha$ and $\beta$ are real valued parameters.
The function $g(t)$ follows the differential equation,
\begin{equation}
 \dot{g}= \alpha g  - i \beta |u_0|^2 u_0.\label{g}
\end{equation}
Adding white noises into \eqref{g}, we obtain a parametric model given by,
\begin{equation}\label{reducedmodel}
\left\{
\begin{aligned}
\dot{u}_0 = & - ic \frac{1}{2} u_0 - id|u_0|^2 u_0 + b f + \beta g \\
\dot{f} = & a f - b^* u_0 + \sigma_1 \dot{W}_f\\
\dot{g} = & \alpha g - i\beta |u_0|^2 u_0 + \sigma_2 \dot{W}_g,
\end{aligned}
\right.
\end{equation}
where we have added an equation of $g$ to represent the second memory term in \eqref{eq: gle-u0}. The problem here is that the nonlinear terms do not conserve energy since we can not control the nonlinear terms in the equation for $g$ unless for $\beta\neq 0$. We suspect that there probably exists different approximations (other than the rational functions) for these kernel functions that give stable parametric models and these are beyond the scope of this paper. Based on this consideration, we do not implement the parametric model in \eqref{reducedmodel} in this paper. 

\section{Pseudo-algorithm for parameter estimation}
This Appendix provides a pseudo-algorithm of the estimation method proposed in \cite{hmm:14}. Consider the following filtering problem,
\begin{eqnarray}
\tilde{x}_j &=& f(\tilde{x}_{j-1}) + \Gamma \epsilon_k, \quad \epsilon_k \sim \mathcal{N}(0,Q),\label{canonical}\\
v_j &=& H \tilde x_j + \epsilon^o_j,\quad \epsilon^o_j \sim\mathcal{N}(0,R), \nonumber
\end{eqnarray}
where $\tilde{x}_j = (x_j,\theta_{d,j})$ denote the augmented state and deterministic parameters. Here, we assume a persistence model for the deterministic parameters, $\theta_{d,j}= \theta_{d,j-1}$. We attempt to estimate $\tilde x_j$ as well as $Q$ and $R$, on-the-fly. Essentially, $Q$ and $R$ are the stochastic parameters through the following relation,
\begin{eqnarray}
Q = \sum_{i=1}^p Q_i \theta_{s,i}, \quad
R = \sum_{i=1}^p R_i \theta_{s,i}.\nonumber
\end{eqnarray}
and our aim is to estimate $\theta_{s,i}$, $i=1,\ldots,p$. For the model in \eqref{sde}, the augmented state-parameters are $\tilde x = (\text{Re}\{ u_0\}, \text{Im}\{ u_0\},\text{Re}\{f\},\text{Im}\{f\}, a_1, a_2, b_1, b_2, c, d)^\top$, the number of stochastic parameters are $p=2$, where $\theta_{s,1}=\sigma_1^2, \;\theta_{s,2}=R$, and
\begin{align}
\Gamma  = \begin{pmatrix} 0 & 0 & \frac{1}{\sqrt{2}} & 0 & 0 & \ldots & 0 \\
 0 & 0 & 0 & \frac{1}{\sqrt{2}} & 0 & \ldots & 0 \end{pmatrix}^\top,\quad  Q_1 = \mathcal{I}_2, \quad Q_2 = R_1 = 0, \quad R_2 = 1.\end{align}

Starting with time index $j=1$, we provide an ensemble of prior statistical estimates, $\{\tilde{x}^{k,-}_j\}_{k=1}^K$, of size $K$ for the primary filter and prior mean $\{\theta_{s,i,j}\}_{i=1}^p$ and covariance $\Theta_j=\mathcal{I}_p$, for the secondary filter. The primary filter for estimating $\tilde{x}_j$ is described in Steps 1-3, while the secondary filter for estimating $\theta_s$ is described in Steps 4-9. 
\begin{enumerate}
\item Apply the ETKF to obtain the analysis ensemble estimate, $\{\tilde{x}^{k,+}_j\}_{k=1}^K$. Let's denote the corresponding Kalman gain and innovation as follows,
\begin{eqnarray}
\tilde{K}_j  &=& P^-_jH^\top (HP^-_jH^\top+ \sum_{i=1}^p R_i \theta_{s,i,j} )^{-1} \nonumber\\
\epsilon_j &=& v_j - H\bar{x}^-_j,\nonumber
\end{eqnarray}
where $\bar{x}^-_j = K^{-1}\sum_{k=1}^K \tilde{x}^{k,-}_j$ denotes the prior ensemble average.  See \cite{hunt:07} for the detail ETKF algorithm.
\item Propagate each ensemble member with the deterministic part of the model in \eqref{canonical} to obtain,
\begin{align}
\tilde{x}^{k,d}_{j+1} = f(\tilde{x}^{k,+}_{j}), \quad k=1,\ldots, K,\nonumber
\end{align}
and form the posterior ensemble by adding a Gaussian noise,
\begin{align}
\tilde{x}^{k,+}_{j+1} = \tilde{x}^{k,d}_{j+1} + \psi^k, \quad \psi^k \sim \mathcal{N}\big(0,\Gamma (\sum_{i=1}^p Q_i \theta_{s,i,j})\Gamma^\top\big), \quad k=1,\ldots, K.\nonumber
\end{align}
\item Define an ensemble approximation for the linear tangent model, 
\begin{eqnarray}
A_j \equiv \nabla f(\bar{x}^+_j)\approx U^d_{j+1}W_j^\dagger,\label{Aj}
\end{eqnarray}
where each column vectors of $U^d_{j+1}$ and $W_j$ are the deterministic forecast ensemble perturbations and the analysis ensemble perturbations, consecutively. In \eqref{Aj}, we denote pseudo-inverse by $\dagger$.
\item Define $K_j = A_j\tilde{K}_j$ and $\phi_j = A_j -K_jH$.
\item For each $i=1,\ldots, p$, construct an observation operator for $\epsilon_j\epsilon_j^\top$, starting with $S_{i,1,0}=0$, let
\begin{eqnarray}
M_{i,j,0} &=&  S_{i,j,0} H^\top,\nonumber\\
F_{i,j,0} &=& HM_{i,j,0} +R_i, \nonumber\\
S_{i,j+1,0} &=& \phi_kS_{i,j,0}\phi_j^\top + \Gamma Q_i \Gamma^\top + K_j R_i K_j^\top.\nonumber
\end{eqnarray} 

\item For each $i=1,\ldots, p$, construct an observation operator for $\epsilon_j \epsilon_{j-\ell}^\top$, where $k>1$. Set
\begin{eqnarray}
M_{i,j,\ell} &=& \phi_{j-1}M_{i,j-1,\ell-1} - K_{j-1}R_i\delta_{\ell,1} \nonumber\\
F_{i,j,\ell} &=& H M_{i,j,\ell}\nonumber
\end{eqnarray}

\item Approximate $\mathbb{E}(v_j v_j^\top)= \sum_{i=1}^p F_{i,j,0} \theta_{s,i,j}$. Suppose if $\epsilon_j = (\epsilon^1_j,\ldots,\epsilon^m_j)^\top$ is $m$-dimensional. Define
\begin{eqnarray}
\sigma_{j,\ell} \equiv vec(\epsilon_j\epsilon_{j-\ell}^\top) = (\epsilon^1_{j}\epsilon^1_{j-\ell}, \epsilon^2_j\epsilon^1_{j-\ell}, \ldots, \epsilon^m_{j}\epsilon^1_{j-\ell}, \ldots, \epsilon^1_{j}\epsilon^m_{j-\ell}, \epsilon^2_j\epsilon^m_{j-\ell}, \ldots, \epsilon^m_{j}\epsilon^m_{j-\ell})^\top.\nonumber
\end{eqnarray}

\item Consider the pseudo observation model for the secondary filter,
\begin{eqnarray}
\sigma_{j,\ell} = \mathcal{F}_{j,\ell}\, \theta_s + \eta_{j,\ell},  \quad \eta_{j,\ell}\sim \mathcal{N}(0,W_{j,\ell}),\quad \ell=1,\ldots, L,\label{pseudo_obs}
\end{eqnarray}
where in our case, $\sigma_{j,\ell} = vec(\epsilon_j\epsilon_{j-\ell}^\top)\in \mathbb{R^+}$, $\mathcal{F}_{j,\ell}=(F_{1,j,\ell},\ldots, F_{p,j,\ell})$, $\theta_s = (\theta_{s,1},\ldots,\theta_{s,p})^\top$, and for each pair of indices $\{k,\ell\}$, construct
\begin{eqnarray}
W_{j,\ell} = \mathbb{E}(\epsilon_j \epsilon_j^\top)\mathbb{E}(\epsilon_{j-\ell}\epsilon_{j-\ell}^\top) + \mathbb{E}(\epsilon_j \epsilon_j^\top)^2\delta_{\ell,0}.\label{W}\nonumber
\end{eqnarray}
Note that $W$ is constructed, assuming Gaussian and independent noises, $\eta_{j,\ell}$. Components of matrix $W$ in \eqref{W} can be rewritten as follows,
\begin{eqnarray}
W^{\alpha,\beta,\gamma,\delta}_{j,l} =   \mathbb{E}(\epsilon^{\alpha}_j \epsilon^{\gamma}_j)\mathbb{E}(\epsilon^{\beta}_{j-\ell}\epsilon^{\delta}_{j-\ell}) + \mathbb{E}(\epsilon^{\alpha}_j \epsilon^{\delta}_j)\mathbb{E}(\epsilon^{\beta}_j\epsilon^{\gamma}_j)\delta_{\ell,0}.\nonumber
\end{eqnarray}
\item Perform a secondary Kalman filter $L$-times to sequentially update $\theta_{s,i,j+1}$ with observation models in \eqref{pseudo_obs} one at the time, assuming that the dynamics of these parameters are persistence, $\dot{\theta}_{s,i}=0$. Now we can repeat Step~1 above for the new assimilation time.
\end{enumerate}

\end{document}